# Towards full-stack functionalist theories of consciousness: Identifying the functional profile

**Sushrut Thorat, Paras Chopra**
{sushrut.thorat, paras}@lossfunk.com

## Abstract

*Theories of consciousness often explain different pieces of the phenomenon, leaving open how to construct an account of the whole. If consciousness is functional, a basic question is what awareness of a content changes in how a system can use it. We review experiments comparing how information X is used in a process Y with and without awareness of X, while constraining differences in the information available about X. The resulting preliminary functional profile links awareness most clearly to attentional control, with narrower evidence for metacognitive evaluation, while simple feature–location binding and retrieval can persist without awareness. But these associations constrain rather than explain consciousness. A full-stack theory must also account for the structure and manner of presentation of perceptual and affective experience, subject–world organisation, awareness judgments, and problem reports. Attention Schema Theory illustrates a possible bridge: using a functional association to motivate an internal process model, then proposing that its information also underlies awareness and associated reports. Generalising this approach, we propose using both functional and explanatory profiles to guide the abductive construction of candidate process architectures, implementing and intervening on them, and progressively testing and refining them against further evidence.*

## Introduction

Understanding consciousness matters not only because experience itself demands explanation, but because judgments about which systems are conscious bear on how we understand other minds, personhood, welfare, and moral status. Our strongest empirical grip on consciousness comes from ordinary adult humans, where reports, behaviour, and knowledge of the human brain can be considered together, even though none provides a complete characterization of consciousness. The problem becomes harder in infants, brain-injured patients, non-human animals, and artificial systems, where some of these sources of information are unavailable, substantially altered, and harder to acquire (Bayne et al., 2024). From a constructivist standpoint, artificial systems now question the substrate-dependence of consciousness because increasingly sophisticated cognitive capacities can now be realised in physical substrates very different from brains. This motivates a basic theoretical question: which properties of a system are actually relevant to consciousness? One attractive place to look is the organisation and interaction of the human consciousness-related mental/bodily-processes, because such descriptions can in principle be compared across very different physical implementations (Butlin et al., 2026). Thereafter, any extrapolation to other systems must be earned through predictive success in humans and a justified account of which organisational features are expected to remain invariant across implementations (Chopra, 2026).

Against this background, we take functionalism as a working hypothesis: if a system has the right functional organisation, it has the relevant conscious states (Putnam, 1967). But what is the right functional organisation, and what must it explain? Saying that consciousness is functional does not yet tell us which processes, arranged in what way, are sufficient for seeing red, feeling pain, knowing that one sees, and wondering why any of this feels like anything at all. We use "full-stack" to characterise this missing ambition. A full-stack functionalist theory of consciousness should specify the phenomenon to be explained, the functional organisation that constitutes it and its process-level implementation, why a system would possess such an organisation, and how that organisation accounts for the structured first-person states, judgments, behaviours, reports, and "problem reports" associated with consciousness (Chalmers, 2018). It should not just identify consciousness with one useful operation — broadcast, attention, integration, metacognition — and leave everything else in a philosophical footnote. This does not assume that a single mechanism must explain the entire stack: the relevant organisation may comprise several interacting processes, and whether these can ultimately be captured within one unified account is itself an open question.

How might this ambition be turned into an actual theory? The Attention Schema Theory (AST) can be read as a useful worked example (Graziano et al., 2020; Schurger & Graziano, 2022). It takes seriously a putative content-specific functional association: awareness of a content appears to matter for controlling attention with respect to that content. It then uses this association to motivate a process architecture that is also intended to account for awareness judgments and reports. Whether this specific account is correct is secondary here. The important move is the bridge it attempts to build: from an empirical functional association, through a process-level hypothesis and a reason for that process to exist, to the wider phenomena associated with consciousness. AST develops this move most concretely for attentional control; a full-stack theory would need comparable process-level accounts of the other phenomena associated with consciousness.

Generalising this move, we distinguish three interdependent tasks towards building a full-stack functionalist theory of consciousness. First, the explananda need to be detailed. This includes characterising the structured perceptual and affective character often grouped under "qualia", the apparent presentation of a world to a subject, awareness judgments, and the peculiar reports people make about experience — including that it seems private, immediately known, irreducible, or difficult to explain (Chalmers, 2018; Kammerer, 2026). Second, the awareness-linked functional associations need to be identified. We need to determine which uses of a content change when that content is conscious rather than unconscious, and which remain unchanged or available without awareness (Niikawa et al., 2022; Michel, 2026). Third, theories need to be constructed that connect the processes underlying these associations to the wider explananda. This step necessarily involves abductive jumps: empirical findings constrain a process architecture, but they do not construct it for us. Figure 1 summarises how the explanatory and functional profiles jointly constrain, and are recursively refined through, the construction of candidate process architectures.

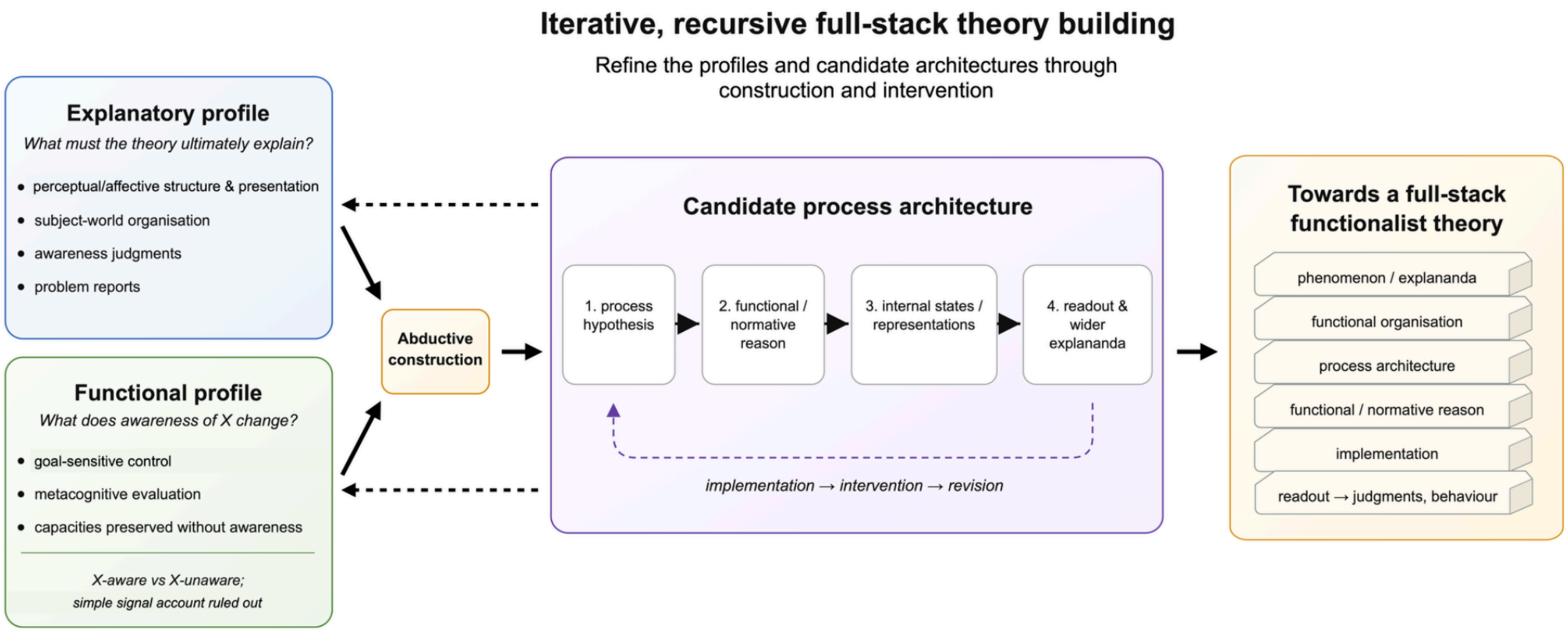

**Figure 1. Framework for constructing a full-stack functionalist theory of consciousness.** *The explanatory profile specifies what the theory must ultimately explain; the functional profile specifies what awareness appears to change in the use of information, alongside capacities preserved without awareness. Together, they constrain, but do not determine, the abductive construction of candidate process architectures. Construction involves linking a process hypothesis to an account of why that organisation exists, what internal states and representations it contains, how it is implemented, and how it relates to the wider explananda, including how readout processes generate judgments, behaviour, and reports. Building and intervening on candidate architectures can in turn refine both the architectures and the profiles.*

The present paper mainly develops the second task. Rather than just cataloguing behavioural or neural differences that accompany X-aware and X-unaware conditions, we ask a content-specific question: for a particular content X and process Y, does awareness of X change whether, or how well, X can be used in process Y?[1] A well-supported absence of such a change is equally informative, because it constrains what awareness of X need not contribute to Y. The aim is therefore to construct a contrastive functional profile of which uses of X change with awareness and which remain available without it. This apparently simple question is difficult to answer cleanly because aware and unaware conditions often differ in more than awareness. In much of the unconscious-processing literature, X is weakened or masked until it is no longer reported, so any corresponding change in Y may reflect reduced signal about X rather than awareness itself (Kouider & Dehaene, 2007; van den Bussche et al., 2013). Conversely, showing that a weakened, unreported X still affects Y establishes a capacity for unconscious processing but does not by itself show that awareness adds nothing. The empirical challenge is therefore to identify awareness-linked changes in the use of X that cannot be reduced to differences in the available information about X (Morales et al., 2015; Michel, 2026).

Several elements of this programme have close precedents. Prior work has made consciousness-related judgments and problem reports explicit explananda (Chalmers, 2018), articulated empirical criteria that theories of consciousness should satisfy (Doerig et al., 2021), distinguished explanatory theories from descriptions of consciousness (Schurger & Graziano, 2022), and examined functional differences between conscious and unconscious processing (Lamme, 2015; Ludwig, 2023; Van Gaal et al., 2012). More recently, Michel (2026) has highlighted the difficulty of attributing such differences to awareness when conscious and unconscious conditions differ in the available evidence. Natural-kind approaches have also proposed iteratively refining indicators of consciousness through patterns of co-occurrence and explanatory inference (Bayne & Shea, 2020; Mckilliam, 2025). At a broader methodological level, Owen & Goodale (2026) question the scientific payoff of grand theories relative to empirical research, while Lopez & Wiese (2025) advocate focusing on simpler theoretical building blocks without committing to particular theories of consciousness. Seth & Bayne (2022) likewise call for computational models addressing both functional and phenomenological properties. Here we develop a particular route towards that aim: identifying content-specific functional constraints through awareness contrasts that address signal-strength explanations, and using these alongside a broader explanatory profile to guide the abductive construction of process architectures. The functional and explanatory profiles provide complementary constraints on theory construction: the former specifies what awareness appears to change in the use of information, while the latter specifies what the resulting organisation must ultimately account for. Refining the explanatory profile can also guide empirical investigations that extend or revise the functional profile. Candidate process architectures can then be generated and progressively exposed to both profiles, rather than beginning from a fixed set of named theories.

[1] Note that we use "awareness" of X to refer to the experimentally indexed conscious status of a particular content, reserving "consciousness" for the broader phenomenon targeted by the theory.

This paper highlights a path towards full-stack functionalist theories of consciousness. We first clarify the wider explananda that such a theory must ultimately address. We then develop the X → Y framework and use it to identify awareness-linked functional associations and capacities preserved without awareness. Finally, we ask how these empirical constraints can guide the abductive construction of process-level theories that connect functional organisation to the broader phenomena of consciousness.

## Constructing the explanatory profile

Before asking what awareness changes functionally, we need a clearer account of what a full-stack theory is ultimately trying to explain. Our starting point is ordinary adult human consciousness, for which we have the richest combination of first-person reports, behaviour, and experimental evidence. Consciousness does not present itself to science as one clean datum. For present purposes, we focus on a cluster of features that recur across experience, judgment, report, and behaviour, without assuming that this profile is exhaustive or that its components correspond to distinct mechanisms (Schurger & Graziano, 2022).

There appears to be a world containing objects, events, relations, and agents. There appears to be a body situated in that world, and a perspective from which parts of it are seen, heard, felt, desired, enjoyed, or suffered. Perceptual and affective states also have characteristic internal structure. Colours, sounds, pains, and other experiences stand in similarity relations that are reflected in behaviour and report: red is more similar to orange than to blue; pitches and other sensory qualities vary along structured dimensions (Fleming & Shea, 2024). A theory should account for why perceptual and affective states have the particular structured character they do.

These states also appear to be organised in a distinctive subject–world form. Objects and events seem to occur in a world around us; bodily states seem to occur here; and perceptual and affective states seem to be present from a particular point of view. This describes an apparent subject–world relation, not a literal inner theatre or an observer inspecting representations. Nevertheless, why cognition is organised and described in this subject–world form is itself part of what a theory should explain (Metzinger, 2007). Beyond these structural relations, a theory must also explain how conscious contents seem to be presented (Chalmers, 2018; Kammerer, 2026). Even knowing that vision is mediated and fallible, we experience objects as simply there before us — not as inner images or inferences we consult (Metzinger, 2007). Likewise, a theory of pain must account for how an episode hurts, how this relates to detecting harm, avoidance and learning, and how its character can be attended to and described. These should be posed as empirical questions: there's no apriori assumption here that experience contains more than can ever be accessed, or that conscious presentation is added by a later introspective or linguistic stage.

Alongside these states are judgments about them. A system can judge that it saw something, felt something, knew something, or failed to be aware of something. It can also generate more puzzling judgments and reports: that experience seems private, immediately present, difficult to communicate completely from a third-person perspective, irreducible to information processing, or somehow left unexplained by a mechanistic account (Chalmers, 2018; Kammerer, 2026). These ordinary awareness judgments and the more unusual "problem reports" need not arise in the same way, but both form part of the explanatory profile considered here.

This profile is intended as a starting point rather than an exhaustive decomposition of consciousness. A successful theory may show that several of these features arise from one organisation, that some are downstream consequences of others, or that distinctions that appear important at the descriptive level dissolve once the underlying processes are understood. A successful account may also explain some apparent features as consequences of how the system represents its own states, rather than as properties those states possess. The point is not to stipulate in advance how many components consciousness contains, but to avoid allowing a theory to explain one part of the profile and silently treat it as the whole phenomenon.

Some aspects of this profile, particularly its putative first-person structure, are not directly available to external observers. Scientific study therefore approaches them through reports, behaviour, and other measurable consequences. Direct reports provide one important route, but we should note they are not transparent recordings of an inner state (Tsuchiya et al., 2015; Fleming, 2020). They depend on categorisation, memory, decision, language, and reporting context, and are therefore transformed readouts of whatever internal information is available to the reporting system. They may omit information that nevertheless influences perception, confidence, action, affect, or bodily control. Relatedly, apparent experiential richness need not imply that all experienced detail is simultaneously available for report. One possibility is that the sense of richness partly reflects a structured representation that can be rapidly inspected and used from shifting task-dependent perspectives. Indirect measures — including similarity judgments, forced-choice behaviour, confidence, error correction, action selection, and physiological responses — can therefore provide complementary constraints on the underlying organisation.

A full-stack functionalist theory must ultimately connect this explanatory profile to an explicit process organisation. It should distinguish the underlying states and processes from the mechanisms that read them out and from the resulting reports and behaviours. It should say what relevant internal states exist, how they are structured, what they represent, what roles they play, and how they become available to different parts of the system. It should then explain how this organisation gives rise to the structured perceptual and affective character and manner of presentation described above, why it takes a subject–world form, and why particular readout processes generate ordinary awareness judgments as well as the more peculiar problem reports. Whether the present profile is complete, and how its elements should ultimately be related, are questions we return to in the Discussion.

## Constructing the functional profile

The explanatory profile specifies what a full-stack theory must ultimately account for, but it does not by itself tell us what functional organisation could account for it. One way to constrain that organisation is to ask where awareness of a content changes how that content can be used (Lamy et al., 2015; Ludwig, 2023). This is not simply the question of whether a consciously perceived content can affect what the system does: in many cases, information about X can influence processing and behaviour even when X is not consciously reported (Kouider & Dehaene, 2007; van den Bussche et al., 2013). The informative comparison is therefore whether awareness changes how that already-registered information can be used. We ask, for a particular content X — an external stimulus or internally generated information — and mental or bodily process Y, whether X is used differently in Y when participants are aware of X than when they are unaware of it, and if so, how. We distinguish Y from the outcome used to measure it: Y is the process-level use of X, whereas the outcome is the behavioural, judgmental, physiological, or other measure through which that use is inferred. The aim is not to catalogue every process that accompanies awareness, but to identify content-specific functional associations, alongside capacities preserved without awareness, that candidate process architectures must explain.

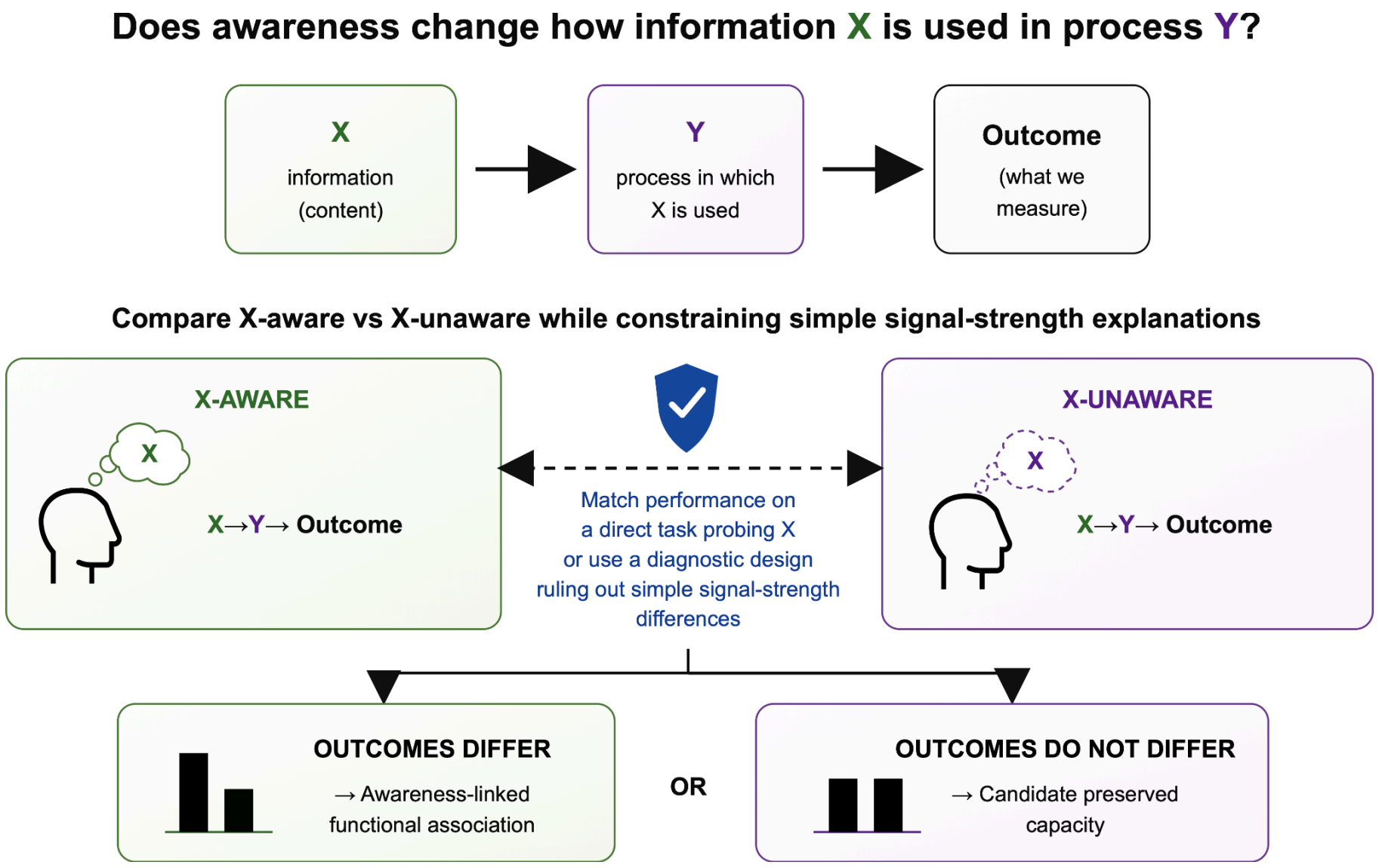

**Figure 2. The X → Y framework for identifying awareness-linked functional associations.** *For a particular content X and process Y, X-aware and X-unaware conditions are compared while constraining poorer information about X as an explanation of any downstream difference, either by matching performance on a direct task probing X or through a diagnostic design. A reliable difference in outcome provides a candidate constraint on the relationship between awareness and the use of X in Y, whereas no detectable difference, alongside positive evidence of the function in the unaware condition, can support the interpretation that the relevant capacity remains available without awareness. Stronger claims require converging evidence across independent studies and paradigms.*

This distinction matters because much of the unconscious-processing literature asks a related but weaker question: whether information that is not consciously reported can still influence behaviour. Such studies provide important evidence that unconscious processing exists, but do not necessarily reveal what awareness contributes. If information about X is weakened in the unconscious condition, differences in Y may reflect differences in available information rather than awareness itself. Conversely, showing that unconscious X can influence Y demonstrates a capability of unconscious processing, but does not show that awareness is functionally irrelevant. A masked stimulus may still prime a response, bias a decision, or influence a later computation; this establishes that some information about the stimulus survived the manipulation, not that the conscious and unconscious versions are functionally equivalent or that awareness adds no further capacity. The question relevant for functionalist theory is therefore not simply whether unconscious information can be used, but whether awareness changes how information is used (Michel, 2026).

We therefore focused on experiments in which awareness of X was explicitly assessed or independently validated for the relevant stimulus regime, and where the design constrained explanations based on differences in the information available about X. Matching performance on a direct task probing X (Morales et al., 2015; Samaha, 2015) and identifying diagnostic dissociations provide different forms of control. Physically identical stimuli need not yield equivalent internal evidence (Charles et al., 2013), while non-monotonic effects or reversals rule against particular signal–outcome relationships rather than all signal-dependent explanations. Each candidate constraint should therefore be interpreted in light of the alternatives its design addresses. These controls are not intended to make aware and unaware states internally identical, or to establish awareness as an independent causal variable. Rather, they constrain explanations in which a functional contrast simply reflects differences in the information available about X. In particular, a reversal or non-monotonic effect can still arise because stronger evidence recruits a control process, such that awareness and the functional change accompany a common transition in processing. Explaining why, and under what conditions, these changes coincide is then a task for process-level theory; their association alone does not establish that they share a mechanism or that this mechanism constitutes consciousness.

We additionally require that Y support a process-level interpretation of how X is used by the system. The relevant question is not whether X produces any measurable trace, but whether the experiment allows us to specify how the system is using X and whether that use changes with awareness. Neural responses, generic priming effects, or the ability to report X itself can therefore be informative without, by themselves, identifying such a process. An awareness-linked neural difference, for example, may constrain the implementation of a candidate mechanism, but unless its computational role is independently understood it does not yet specify what the system is doing differently with X. This is a methodological criterion, not a claim that excluded effects are biologically unimportant: some may reflect important processes whose roles are not yet understood. The same criterion makes capacities preserved without awareness informative: a process that appears similarly available without awareness constrains theory by indicating what awareness may not be required for. The goal is therefore not to accumulate a list of “functions of consciousness”, but to construct a contrastive functional profile of awareness-linked differences and capacities preserved without awareness. Figure 2 summarises this X → Y comparison and the two kinds of functional constraints it can yield.

Applying these criteria during a literature sweep yielded a sparse but informative functional profile. We began from foundational reviews and candidate studies relevant to functional accounts of consciousness, and expanded the set manually through citation-based exploration (Van Boxtel et al., 2010; Lamme, 2015; Graziano et al., 2020; Doerig et al., 2021; Fleming & Michel, 2026; amongst the other references in this paper). Candidate studies were then evaluated according to whether they tested how awareness of a particular content X changed how that content was used in process Y. The search proceeded iteratively until further citation-based exploration ceased to yield additional studies satisfying these criteria. The resulting set should not be interpreted as an exhaustive map of awareness-linked functional associations. Individual experiments should be treated as candidate constraints; stronger claims about an awareness-linked functional association require converging evidence across independent studies and paradigms. The present set therefore identifies cases where awareness appears to alter the use of information, alongside cases where information remains usable without awareness. Figure 3 summarises the three primary cases and the contrastive functional profile they jointly motivate.

## Awareness and attentional control

A first class of findings associates awareness with greater attentional control over information that can influence behaviour.

Tsushima et al., (2006) provide a particularly informative example (Figure 3, left). Participants performed a demanding central rapid serial visual presentation (RSVP) task in which they identified two digits embedded in a stream of letters, while a task-irrelevant random-dot display moved in the background. The proportion of coherently moving dots varied from trial to trial. With no coherent motion, RSVP performance was relatively high. Surprisingly, performance dropped sharply at 5% motion coherence, but recovered as coherence increased: by 20% coherence, the considerably stronger motion signal no longer produced a detectable cost relative to 0%. The effect was therefore non-monotonic. A weak motion signal interfered with the central task more than a stronger one.

Critically, the performance dip occurred around the perceptual threshold for the motion signal. Motion-direction discrimination was measured independently, placing 5% coherence below the threshold for above-chance direction discrimination, whereas the higher coherence levels were discriminable. A closely related 5% coherence regime had also been examined directly in an earlier study from the same group using a closely matched random-dot display: participants did not report detecting coherent motion at this level, and direction-identification and detection performance remained at chance when these judgments were made during exposure (Watanabe et al., 2001). The link to perceptual visibility was further tested within Tsushima et al. by lowering the luminance contrast of the dots. This made coherent motion harder to discriminate, shifting the perceptual threshold towards higher coherence; importantly, the RSVP performance dip shifted with it. Thus, maximal interference was not tied to a particular physical coherence value. It tracked the regime in which the motion signal was strong enough to influence visual processing but remained below the level at which its direction could be reliably discriminated.

The important contrast is therefore not between an effective unconscious signal and an ineffective conscious one. At low coherence, task-irrelevant motion strongly disrupted behaviour; at higher, discriminable coherence, the system was better able to prevent the same kind of information from interfering with the task. Tsushima et al. interpreted this as a difference in inhibitory control. Consistent with this interpretation, fMRI responses to task-irrelevant motion showed maximal activity in motion-sensitive area MT+ around the 5% condition, while lateral prefrontal cortex (LPFC) activity remained low; at higher coherence, LPFC activity increased while MT+ activity was reduced. They proposed that weak motion signals can activate sensory representations without sufficiently engaging prefrontal control, whereas stronger, perceptually available signals recruit control that suppresses their distracting influence. The neural account is a mechanistic interpretation rather than proof, but the behavioural result itself provides the relevant constraint: awareness of a distractor is associated not with whether it can influence processing, but with whether its influence can be brought under control.

Additional studies provide qualified support for this broader pattern, while leaving particular awareness, information-strength, or functional-interpretation questions unresolved. Lin & Murray (2015) found that, when cues repeatedly occupied one location, invisible cues still facilitated responses there whereas visible cues did not. In subsequent cue-absent blocks, the visible-cue group showed a bias against the previously cued location, consistent with suppression. When cue locations varied, both visible and invisible cues facilitated responses. This provides qualified support for awareness-linked regulation of repeated attentional influence. However, the masked group was selected using nonsignificant performance on a separate localization test, leaving uncertainty about residual awareness; visible and masked cues also differed physically.

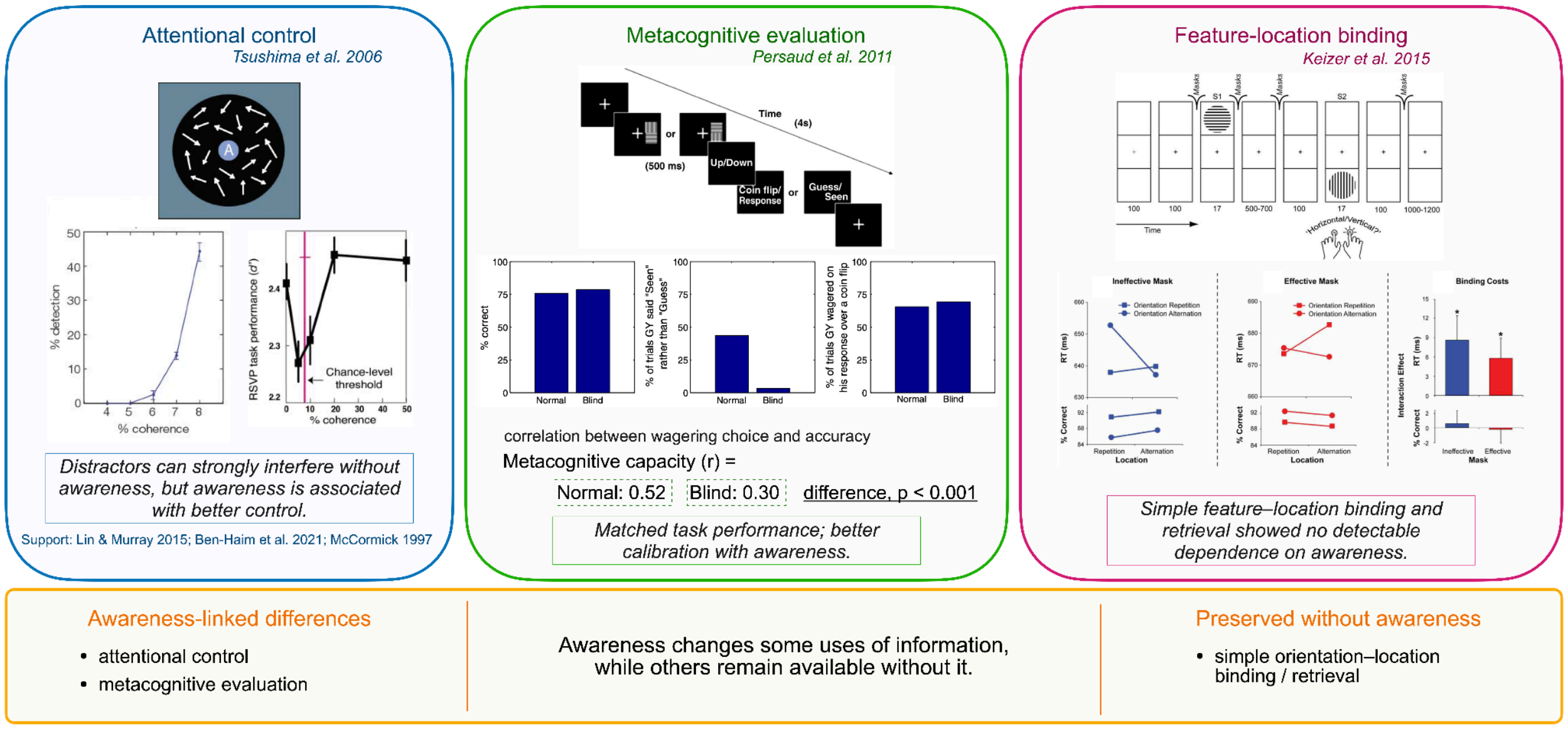

**Figure 3. Awareness-linked differences in the use of information.** *Three primary cases illustrate the current functional profile.* **Left:** *in Tsushima et al. (2006), task-irrelevant motion below but near the separately measured motion-direction threshold produced the largest disruption of a rapid serial visual presentation (RSVP) task, whereas stronger, discriminable motion produced less disruption; Watanabe et al. (2001) independently found a closely matched 5% motion regime to be undetectable.* **Middle:** *in blindsight patient GY, discrimination performance was matched between the normal and blind visual fields, but wagering (confidence) was better calibrated to accuracy in the normal field, associating awareness with better metacognitive evaluation (Persaud et al., 2011).* **Right:** *simple orientation–location binding and retrieval remained detectable when the first stimulus was effectively masked, and the binding effect did not differ detectably between effective and ineffective masking (Keizer et al., 2015). Together, these findings suggest that awareness changes some uses of information while others remain available without it. Stimulus and task schematics and plots were adapted from the cited studies.*

Ben-Haim et al. (2021) provide a related case of flexible rule-governed control. In the fixed-duration session of Experiment 7, cues were presented for 17 ms throughout and participants were instructed from the outset to respond to the location opposite the cue. Trials rated as unaware remained below chance, consistent with automatic cue-driven responding, whereas higher-awareness trials were above chance, indicating successful use of the opposing rule. However, internal cue-location information was not matched across awareness reports. Stronger cue information could itself enable more successful use of the opposite-location rule, so we treat this as supporting evidence for awareness-linked flexible control rather than a primary constraint.

McCormick (1997) provides further supporting evidence for strategic attentional control. Participants knew that targets usually appeared opposite a peripheral cue. In Experiment 2, target identification was faster opposite the cue when participants reported seeing it, but at the cue's location when they did not. A direct interaction supported this difference between awareness conditions. However, usable cue-location information was neither measured nor matched: stronger information about the cue could itself enable more successful orienting towards the expected target location. We therefore treat this as supporting evidence for awareness-linked attentional control rather than a primary constraint.

Here, one example of a promising study, which does not qualify under our criteria, is Persaud & Cowey (2008) which provides a related case in blindsight patient GY. In an exclusion task, GY could follow the instruction to report the location opposite a grating in his normal visual field, but in his blind field he tended instead to report the grating's actual location, and this tendency increased with grating contrast. However, although stimulus contrasts were chosen to equate task difficulty, accuracy when reporting the grating's actual location remained substantially higher in the normal field than in the blind field (95% versus 64%). Inclusion performance was not measured at the lower contrasts, preventing a comparison of exclusion under demonstrated performance matching. The contrast-dependent increase in blind-field errors therefore does not establish that equally informative conscious evidence would be used differently. Contrast levels were also tested in blocks, limiting interpretation of that trend. We therefore discuss this result as relevant context rather than evidence for the present functional profile.

Together, these studies suggest a broader pattern: without awareness, X can still exert stimulus-driven influence; with awareness, that influence can become more regulable via attentional capture.

## Awareness and metacognitive evaluation

A second class of findings concerns metacognition: how well the system can evaluate the accuracy of its own perceptual decisions. Persaud et al. (2011) studied the well-characterised blindsight patient GY (Figure 3, middle), whose left primary visual cortex is extensively damaged. Stationary stimuli presented in his blind right visual field can support above-chance discrimination despite very little reported visual awareness. GY judged whether a vertical grating appeared above or below a horizontal grating, with the stimuli presented either in his normal or blind visual field. Crucially, stimulus contrast was independently titrated across the two fields so that discrimination performance was matched: GY was 76% correct in the normal field and 79% correct in the blind field. Achieving this required dramatically weaker stimuli in the normal field than in the blind field. Despite comparable discrimination, in separate interleaved blocks GY reported seeing the stimuli on 43% of normal-field trials but only 3% of blind-field trials.

In the wagering blocks, GY made a post-decision wager after each discrimination. He could either stake money on his perceptual response being correct or take a 50:50 coin flip; choosing to wager on his response therefore indexed confidence in that decision. Importantly, his overall willingness to wager was virtually identical across the two fields — 66% versus 69% — so the critical difference was not simply that he was less confident in blindsight. Rather, his confidence was better calibrated in the normal field: wagering distinguished correct from incorrect decisions more strongly there than in the blind field (correlation between wagering and accuracy: $r = .52$ versus $r = .30$). The same numerical pattern appeared in a contrast-restricted comparison in which objective performance was higher in the blind field, while metacognitive calibration remained higher in the normal field (.45 versus .25). Thus, comparable perceptual discrimination performance could coexist with a substantial difference in the system's ability to evaluate whether those decisions were correct. This provides a distinct functional constraint: awareness of X is associated with better metacognitive access to, or evaluation of, the system's processing of X.

Here, one example of a promising study, which does not qualify under our criteria, is Vlassova et al. (2014), which reports a related finding. Adding suppressed motion evidence improved first-order decision accuracy without a detectable increase in metacognitive sensitivity. Thus, the additional suppressed evidence benefited the first-order decision without a corresponding metacognitive benefit, but the design does not directly compare the use of matched evidence when conscious versus unconscious. Moreover, because the additional suppressed evidence followed the visible evidence, differential temporal weighting of evidence for choice and confidence could contribute to the dissociation (Zylberberg et al., 2012). We therefore discuss this result as relevant context rather than evidence for the present functional profile.

## A candidate negative constraint: feature–location binding

A useful candidate for negative constraint comes from Keizer et al. (2015) (Figure 3, right), who asked whether awareness is necessary for a simple form of visual feature binding. Participants saw two successive Gabor patches, S1 and S2, each characterised by an orientation and a location. They responded only to the orientation of the visible S2. The logic was that if S1's orientation and location had been bound together into a temporary representation, encountering one of those features again on S2 should retrieve the other. Consequently, partial repetitions — for example, repeating S1's location while changing its orientation — should create conflict and slow the response relative to conditions in which both features repeat or both change. This interaction between orientation and location is the study's behavioural signature of binding.

Awareness of S1 was manipulated using effective versus ineffective masking. Mask strength was individually calibrated to drive detection of effectively masked S1 to chance, and visibility was evaluated again after the binding task; participants showing reliable above-chance detection were excluded. Despite this difference in visibility, the orientation–location binding effect was not detectably reduced by effective masking. In the primary analysis, which excluded participants with poor target-task performance, binding costs were 8.6 ms with ineffective masking and 5.8 ms with effective masking. Both were significant in one-sided tests, with no significant interaction between masking and the binding effect. A more inclusive analysis retained an overall binding effect and again found no masking-by-binding interaction, although the binding costs within the individual masking conditions were no longer significant. The strongest conclusion is therefore not that all feature binding is entirely identical with and without awareness, but that this simple orientation–location binding operation showed no detectable dependence on awareness. On this interpretation, S1’s orientation and location are bound together, retained until S2, and automatically retrieved when a feature repeats. This places a useful boundary on the positive findings above: whatever awareness adds to attentional control and metacognitive evaluation, at least some short-lived feature binding and retrieval can proceed without it.

## Further supporting findings

Further studies suggest that awareness may be associated with how information is used in several downstream cognitive processes. These include immediate correction and subsequent adaptation following errors, interference from previously presented visual cues, and semantic facilitation between successive stimuli. Although these findings broaden the candidate functional associations of awareness, they provide qualified support because differences in internal information remain unresolved and some behavioural effects admit alternative process-level explanations.

Klein et al. (2007) examined errors in a task requiring participants to look away from a peripheral target. Eye tracking recorded initial movements towards the target and subsequent corrections; participants then reported whether their initial movement had been correct. Immediate corrections were more frequent after unreported errors, whereas next-trial accuracy was higher after acknowledged errors. This suggests different relationships between awareness and immediate correction versus subsequent adaptation. However, acknowledged errors involved larger eye movements, leaving error information unmatched, and retrospective reports could not establish whether awareness was absent during correction. We therefore treat these findings as qualified support for error-guided behavioural adjustment.

Dali & Hester (2022) examined whether awareness of an error predicted better subsequent response inhibition. Participants pressed a button for each letter but withheld their response when a letter immediately repeated. Participants knew that, after an error, the next withholding trial would usually involve the same letter, giving participants an opportunity to avoid repeating the mistake. Withholding when that letter returned was more successful after acknowledged than unreported errors. Dali et al. (2023) additionally showed that this awareness advantage was greater when the failed letter recurred than when a different letter appeared—a comparison the earlier study had not directly tested. However, neither study matched internal information about the error or retention of the failed letter. Stronger retained information could itself explain the later advantage. Acknowledging an error also required a distinct button press, which could reinforce processing of the event. Both studies therefore provide supporting evidence for error-guided adaptation while leaving these information-strength and reporting alternatives unresolved.

Lamy et al., (2015) examined whether awareness of a preceding colour cue changes its interference with a subsequent target. Participants reported the orientation of a colour-defined target and then whether they had seen the cue. In the first two experiments, an irrelevant-colour cue slowed responses when it appeared at the target's location rather than elsewhere, but only when the cue was reported as seen. The authors interpreted this same-location cost as a cost of updating an object's earlier representation when its features change. In a third experiment, both cue colours were potentially relevant; colour-mismatching cues produced less facilitation at the cued location when seen than unseen, although this difference was marginal. Internal cue information was not matched, and the response-time effects do not uniquely identify object updating. Moreover, Keizer et al., (2015) found interference from previously bound features without awareness of the initial stimulus, challenging a general claim that awareness is necessary to form and retrieve such representations. We therefore treat Lamy et al.'s findings as qualified support for awareness-linked colour-mismatch interference, while leaving its mechanism and relationship to other binding effects unresolved.

Murphy & Bloom (2015) examined semantic facilitation during the attentional blink, in which identifying one target makes a second, shortly following target difficult to report. Participants viewed a rapid word stream containing three coloured targets: the second appeared during the blink, while the third appeared later. All three were reported after the stream. Semantic relatedness between the second and third words benefited identification of the third more when the second was correctly reported than when it was missed, with direct interactions in both experiments. However, although presentation settings were fixed within each experiment, internal evidence was not matched, and delayed report could reflect encoding or retention as well as awareness during use. Shapiro et al. (1997) had also found facilitation from missed second words, cautioning against a general claim that awareness is necessary. We therefore treat Murphy and Bloom's result as qualified supporting evidence for an awareness-linked difference in semantic facilitation.

## The emerging functional profile

Taken together, these studies suggest a selective functional profile for awareness. The strongest evidence links awareness to attentional control and metacognitive evaluation. Further findings suggest associations with error-guided adaptation, visual cue–target interference, and semantic facilitation, although these remain less secure. Conversely, simple feature–location binding and retrieval can occur without awareness.

The evidential strength is uneven: attentional control has several supporting findings of varying strength; metacognitive evaluation rests mainly on one matched-performance study; and feature binding on one paradigm. The additional associations remain qualified by information-strength, process-level, and reporting alternatives. Nevertheless, this emerging profile provides a promising starting point for theory construction: a concrete set of awareness-linked differences and preserved capacities that a candidate architecture must explain. The next theoretical challenge is abductive: what process organisation could explain this pattern while also accounting for the broader phenomena of consciousness?

## From empirical constraints to process architecture: abductive construction

The functional profile constrains but under-determines process architecture. Moving from associations to mechanisms therefore requires abduction: proposing architectures that would make the observed pattern intelligible, then constructing and testing them against further constraints (Magnani, 2009; Dellsén, 2024).

### A worked example: Attention Schema Theory

A useful way to begin abductive theory construction is to start with one well-supported functional association and ask what process architecture would make it intelligible. The Attention Schema Theory provides a useful worked example of this move. Start with the empirical association highlighted above: attention can be drawn towards information without awareness, while awareness is associated with better control over that attention (Graziano & Webb, 2015). AST asks what process could explain this particular pattern. Graziano appeals to a general idea from control theory: controlling a complex and changing process can benefit from an internal model of that process (Conant & Ashby, 1970; Graziano et al., 2020). Attention is itself something that must continually be directed, maintained, and redirected. AST therefore proposes that the brain constructs a simplified model of its own attention — an attention schema — which tracks relevant properties of attention and helps predict and control its deployment. On this account, attention to X can remain when awareness of X is absent, but control over that attention is compromised because the system lacks an adequate model of its attentional state with respect to X. This provides a process-level explanation for the awareness-linked attentional-control association rather than merely redescribing it. Graziano explicitly develops this analogy with body schemas and argues that awareness should normally track attention while becoming dissociable from it in precisely those cases where control is impaired.

But this only gets AST part of the way towards a theory of consciousness. A second abductive move identifies the information contained in the attention schema with the information underlying judgments of awareness. The system does not merely deploy attention; it also represents itself as standing in a simplified relation of awareness to the attended content. The attention schema is therefore proposed to play two roles at once: helping control attention and supplying the internal information on which the system bases claims such as “I am aware of X”. This is the central bridge from a functionally motivated process architecture to the awareness-related part of the explanatory profile.

AST then makes a further move towards the broader explananda. Because the schema is schematic, it does not represent the detailed physical machinery through which attention is implemented. Instead, the system represents itself as possessing a relatively simple relation of awareness to a content. Graziano argues that this impoverished self-description helps explain why awareness is reported as something vague, private, and apparently non-physical rather than as a complicated neural computation. In this way,

AST attempts to connect an empirically motivated function, a computational reason for a process to exist, an internal representation, and the peculiar reports people make about consciousness.

This sequence is exactly why AST is useful as a template for full-stack theory construction (Schurger & Graziano, 2022). It does not stop at "awareness is associated with attentional control". It asks why that association might exist, proposes a process architecture, and then tries to use the information generated by that architecture to explain awareness judgments and problem reports. At the same time, the later parts of this chain are much less developed than the first. Showing why a model of attention would aid attentional control is not yet the same as showing why the information in such a model should generate the full explanatory profile outlined earlier. In particular, AST as currently developed does not yet specify how the attention-schema mechanism, together with the wider cognitive architecture, yields the structured character and manner of presentation of perceptual and affective states — why a world seems directly encountered, or a bodily state hurts — the apparent subject–world organisation of experience, or the specific forms of immediacy and irreducibility expressed in problem reports. Chalmers and, more recently, Kammerer make closely related criticisms: a simplified model may help explain why awareness is represented incompletely, but incompleteness alone does not yet explain why experience seems so peculiarly irreducible or why these intuitions are so compelling (Chalmers, 2020; Kammerer, 2026).

This gap is important for our framework. The process architecture should not merely reproduce one awareness-linked functional association and then append a verbal story about qualia. The same architecture, or principled extensions of it, must eventually connect back to the explanatory profile with which we began. What information does the architecture make available? How is that information structured? Why does it produce a subject–world form? Why do particular readout processes generate ordinary awareness judgments and reports as well as problem reports? These are additional constraints, not decorative philosophical questions to be dealt with after the cognitive machinery has been specified.

Computational construction can help sharpen these abductive moves. Existing implementations of AST have begun to ask whether an internal representation of attention actually improves attentional control, what information such a representation needs to contain, and whether it can emerge through learning rather than being stipulated in advance (Wilterson & Graziano, 2021; Piefke et al., 2024). Such models do not establish AST as a theory of consciousness, but they force a verbal proposal to make choices that can be tested: what exactly is represented, when is the representation useful, and what breaks when it is perturbed. In recent work, for example, an attention-related representation emerged when keeping track of the attentional state was useful for control, and the representation did not need to be a perfect copy of that state (Piefke et al., 2024). The broader point is not the particular model, but what construction can do to theory: implementing a proposed process can reveal which parts of the original verbal hypothesis were necessary, which were incidental, and which new assumptions are required (Guest & Martin, 2021).

In this sense, model construction can become part of the abductive process itself. One can propose a process architecture, build it, manipulate its components, observe where the intended functional profile does or does not emerge, and revise the hypothesis accordingly. This resembles a manipulative form of abduction (Magnani, 2009): intervention and construction do not merely test a finished theory, but help generate and refine the theory. For a programme committed to process-level explanations, this seems especially useful. The holes in a proposed functional organisation become considerably harder to hide once one tries to make the relevant states and transitions actually work together.

## Expanding the constraint set

A process architecture motivated by one functional association must then be exposed to the rest of the functional profile. If an attention schema, or some related process, helps explain the association between awareness and attentional control, does the same machinery — perhaps after a principled extension — also help explain why awareness is associated with better metacognitive calibration? At the same time, the architecture need not explain feature–location binding itself, but it should permit this operation to remain available when awareness of X is absent. Each additional finding therefore places pressure on the same candidate process architecture rather than automatically licensing another mechanism.

The broader explanatory profile should also guide construction from the outset. A tractable strategy is to begin with a small number of strong empirical constraints, while asking which candidate architectures could also address the experiential phenomena under investigation. Construction and intervention can then refine both profiles and their proposed connections. The explanatory profile is not merely a final test of an otherwise completed functional architecture.

This also changes how existing theories of consciousness enter the programme. We need not begin by treating the existing landscape as a fixed set of alternatives and asking which named theory best fits the observations. Existing theories can instead provide useful process hypotheses that may be retained, modified, combined, or discarded as constraints accumulate. Together, the functional and explanatory profiles define the observations and phenomena a theory must account for. Abductive theory construction then generates a "candidate lot" of process architectures. Comparisons among candidates become informative only once they are developed far enough to face the same constraints.

The present functional profile is therefore a starting point. The immediate theoretical task is to construct process architectures that make its awareness-linked associations and preserved capacities intelligible, develop those architectures against the wider explananda, and use implementation and intervention to refine both architectures and profiles.

# Discussion

The present work offers a starting point for constructing full-stack functionalist theories of consciousness. Our empirical investigation identifies a small but informative set of awareness-linked functional associations and capacities preserved without awareness. The clearest positive evidence concerns attentional control (Tsushima et al., 2006) and metacognitive evaluation (Persaud et al., 2011), with additional, qualified support for attentional control from related paradigms (Lin & Murray, 2015; Ben-Haim et al., 2021). Further studies suggest possible associations with error-guided adaptation, visual cue–target interference, and semantic facilitation. Conversely, simple orientation–location binding and retrieval remained detectable without awareness in one experimental paradigm (Keizer et al., 2015). These findings are not uniformly supported: attentional control has several supporting studies of varying strength, metacognitive evaluation rests primarily on a particularly informative matched-performance comparison, and the binding constraint currently rests on a single paradigm. Nevertheless, the resulting profile provides concrete empirical constraints for theory construction, while leaving open what further processes might prove relevant. The apparent sparsity of this profile also need not imply that awareness has little functional importance. Much of the available evidence comes from brief, controlled experiments (Lamme, 2015), whereas differences in control, reliability, or flexibility may accumulate across repeated, temporally extended behaviour. Awareness could therefore have a relatively thin instantaneous functional footprint but substantial consequences over time.

Importantly, these findings do not establish attentional control and metacognitive evaluation as “the functions of consciousness”. An association between awareness of an entity X (e.g., sensory inputs or thoughts) and a change in how X is used does not demonstrate that awareness causes that change, much less that the process involved constitutes consciousness. Both may arise from a common underlying organisation. Conversely, the preservation of a capacity without awareness helps identify operations that need not depend on whatever additional organisation a candidate theory proposes. The functional profile thus serves a different purpose from identifying a single defining operation: it specifies a pattern of differences and preserved capacities that candidate architectures should explain.

One intriguing possibility is that information about entity X can already be represented and used by the system, while additional machinery models properties of the system's own processing of X. Attention Schema Theory (AST) develops this idea for attention, proposing that an internal model of attentional state facilitates attentional control (Graziano et al., 2020). Metacognitive and perceptual reality-monitoring accounts offer related possibilities for representations concerned with the reliability or status of perceptual processing (Fleming, 2020; Lau, 2022). Could these reflect different components of a more general architecture for modelling and controlling internal processes? Such an organisation might help explain why awareness is associated with attentional control and metacognitive evaluation, and perhaps some forms of error-guided adaptation, while allowing simpler operations such as feature binding to proceed without awareness. Whether these capacities rely on a common process, several coupled processes, or substantially different architectures remains an open empirical and computational question. The present functional profile motivates exploring these possibilities; it does not decide among them.

However, identifying a process architecture that reproduces these functional associations is only part of the explanatory task. AST illustrates two distinct abductive moves. The first proceeds from the association between awareness and attentional control to a functionally motivated internal model of attention. Computational implementations can test whether such a model aids control, what information it contains, and whether it can emerge through learning (Wilterson & Graziano, 2021; Piefke et al., 2024). The second move proposes that the model’s information also underlies awareness judgments and reports: the schema represents the system as attending to X, but omits the physical machinery of attention, helping explain why awareness is described as a seemingly simple, non-physical relation to X (Graziano et al., 2020). This bridges a functionally useful representation to the wider phenomena of consciousness. But the functional utility of the schema does not, by itself, establish this further explanatory role. In particular, why should such a representation account for the structured, first-person character of experience, its apparent immediacy, or the specific intuitions of irreducibility associated with it (Chalmers, 2020; Kammerer, 2026)? The second abductive step therefore requires independent explanatory development and evidential support. The same challenge applies to other architectures motivated by functional associations.

Here the explanatory profile becomes essential. The structure and manner of presentation of perceptual and affective states, their apparent subject–world organisation, ordinary awareness judgments, and problem reports are not secondary features to be explained after the functional machinery is complete. They are constraints on its construction. For example, a theory must account for the structured relations among perceptual qualities, why objects appear present in a world rather than as representations being consulted, and what distinguishes hurting from merely detecting and responding to harm. It must also explain why experiences are described as private, immediately known, or difficult to reconcile with mechanistic descriptions (Chalmers, 2018; Kammerer, 2026). These features need not arise from distinct mechanisms: some may share an underlying origin, others may be downstream consequences of how the system represents or evaluates its own states, and some may vary across development, culture, or context. Determining these relationships is part of the theoretical problem, rather than something to stipulate in advance. Candidate architectures may also differ in whether their awareness-related processes merely track perceptual and affective states or participate in the organisation constituting their conscious presentation.

The explanatory profile itself also needs refinement. Although putative first-person structure is not directly observable by others, regularities across reports, judgments, behaviour, and physiology can provide empirical constraints. Perceptual similarity judgments and quality-space analyses offer one route towards characterising experiential structure (Fleming & Shea, 2024). More detailed first-person methods, including microphenomenological interviewing, may reveal dimensions and temporal organisation that coarse awareness ratings miss (Petitmengin et al., 2019; Berkovich-Ohana et al., 2020). However, such methods also raise an important question: might they preferentially capture relatively "thick" experiences, involving explicit monitoring and greater readiness for report, rather than the more ubiquitous, relatively "thin", less explicitly monitored experiences of everyday life (Schooler, 2002; Heimann et al., 2023)? Similar concerns arise when studying altered states, where both experience and its monitoring may change. We use "thin" and "thick" provisionally to distinguish degrees of explicit monitoring, rather than the richness of experience or the detail of its description. Apparent richness, for instance, might partly arise from a structured representation supporting rapidly shifting, task-dependent inspection, without every detail being simultaneously available for report. The challenge is to distinguish changes in experience from changes in access to, evaluation of, or description of experience, rather than assuming that experience is produced by a later introspective or linguistic stage. Refining these distinctions can guide theory construction, while attempts at construction may reveal where the explananda remain underspecified.

What does this programme imply for the hard problem of consciousness? We do not assume that explaining the relevant functional organisation must leave an irreducible explanatory residue, nor that a process-level explanation automatically dissolves the problem (Chalmers, 1995). A useful historical comparison is the question "What is life?". Biological phenomena once motivated appeals to a special vital principle, whereas progressively detailed explanations of metabolism, reproduction, regulation, and other processes transformed much of that mystery into tractable scientific questions. Klein & Barron (2020) draw on this history to argue that scientific progress on consciousness may likewise involve decomposing an apparently singular mystery into increasingly precise explanatory targets. The analogy is suggestive rather than decisive: consciousness may differ from life precisely because phenomenal experience appears to present a further explanandum beyond the functions associated with it (Chalmers, 2003). Nevertheless, we see value in pursuing an explicit process-level account before deciding what remains unexplained. Such an account may progressively deflate aspects of the apparent mystery by explaining both the organisation underlying experience and why that organisation gives rise to the characteristic judgments and reports about it. How far this strategy succeeds is a question for the research programme, not a conclusion assumed at its outset.

Constructing candidate architectures should also help make these explanatory commitments more precise. Implementing a verbal proposal forces decisions about what information is represented, how internal states interact, and which components are necessary for particular capacities (Guest & Martin, 2021). Intervention and ablation can then reveal whether the proposed organisation actually produces the intended functional profile, whether its assumptions are essential, and what further mechanisms are required. Construction can therefore contribute to the abductive process itself, generating and refining hypotheses rather than merely testing already completed theories (Magnani, 2009). This is particularly important for the proposed bridge to the explanatory profile: specifying how internal representations are structured, accessed, and transformed into judgments and reports makes the relevant claims more susceptible to empirical investigation.

The biological implementation of candidate architectures provides another source of constraints. We have deliberately not treated awareness-linked neural signatures as functional associations unless the computation they implement can be independently characterised. A neural difference between aware and unaware conditions does not, by itself, establish how the system uses the relevant information differently (Krakauer et al., 2017; Schurger & Graziano, 2022). Nevertheless, neural anatomy, connectivity, temporal dynamics, lesion effects, and causal perturbations should constrain which proposed architectures could plausibly operate in humans (Kaplan & Craver, 2011). Functional and neural evidence thus contribute at different stages of the programme: the former helps identify what processes must be explained, while the latter helps constrain how proposed processes could be implemented. The challenge is to use biological evidence without assuming that every contingent feature of the human implementation is constitutive of consciousness across possible systems.

Empirical support for a candidate architecture should be earned progressively. An initial achievement is explanatory unification: accounting for several observations through a coherent organisation rather than assigning an independent mechanism to each. Yet this is partly accommodative when those observations motivated construction in the first place. Stronger support arises when the architecture explains observations that did not shape its development (termed “use-novel evidence”; Negro, 2024) and stronger still when it generates prospective predictions that distinguish it from plausible alternatives. These tests might involve additional functional associations, finer-grained experiential phenomena, neural perturbations, or populations in which particular processes are selectively altered. Predictions should follow as directly as possible from the candidate's core commitments rather than from auxiliary assumptions added to accommodate each finding, a recurring difficulty in evaluating consciousness theories (Yaron et al., 2022). Critically, support for one link in the explanatory chain must not automatically be inherited by later links. An architecture may successfully explain attentional control while remaining inadequate as an account of conscious experience.

The empirical foundations of this programme also require continued scrutiny. Aware and unaware conditions frequently differ in the available information about X, and even physically identical stimuli need not produce matched internal evidence (Charles et al., 2013; Michel, 2026). Behavioural differences may therefore reflect information strength or other processing transitions rather than awareness itself. Awareness reports also depend on additional cognitive operations (Fleming, 2020; Tsuchiya et al., 2015), while nonsignificant direct discrimination alongside a significant indirect effect does not by itself establish a genuine unconscious-processing advantage (Meyen et al., 2021). Conversely, the absence of a detectable interaction does not establish functional equivalence. These concerns do not undermine the value of the functional profile, but specify the empirical standards needed to strengthen and expand it. Furthermore, there is no exhaustive inventory of processes Y or reason to expect existing experiments to have sampled them uniformly. The current profile should therefore be treated as a developing set of constraints, not a complete account of awareness-related capacities.

Testing candidate architectures beyond the paradigms and populations that motivated them offers important opportunities for independent validation. Clinical and developmental differences, altered states, and temporally extended, naturalistic behaviour may reveal dissociations or regularities obscured by conventional experiments. Such settings can test whether an architecture generalises, where its boundaries lie, and which of its assumptions survive new constraints. Animals and artificial systems offer further opportunities to examine the proposed organisation across different implementations, provided that the relevant functional invariances are justified (Chopra, 2026). While predictions about consciousness in these systems cannot independently validate consciousness attributions, their behaviour and internal organisation can still test specific claims of the architecture. Together, these extensions can help distinguish candidate theories and progressively establish their explanatory scope.

This approach also offers a way to reconsider how theories of consciousness are constructed and compared. Existing theories frequently address partially overlapping explananda and draw on different subsets of evidence, making direct comparison difficult. Our framework seeks to establish common explanatory ground: the functional and explanatory profiles, alongside implementation constraints, specify the observations and phenomena requiring explanation — the “explanatory lot”. Abductive construction then generates the “candidate lot” of process architectures that might account for them. Existing theories can supply hypotheses and components that are retained, modified, combined, or discarded, rather than constituting a fixed set of alternatives. Comparisons become more informative once candidates are developed sufficiently to confront the same constraints. The “bad-lot problem” remains: the best available candidate may still be wrong, or the correct architecture may not yet have been considered (Dellsén, 2024). Nevertheless, progressively refining the explanatory lot while constructing and testing new candidates offers a path towards a cumulative science of consciousness, in which theory development is guided by increasingly precise and shared empirical constraints.

## A research agenda

The framework suggests several concrete directions for further investigation:

1. *Study awareness in temporally extended and naturalistic behaviour.* Develop tasks requiring sustained attentional control, repeated error correction, updating, and learning. Test whether differences observed in isolated trials accumulate, diminish, or interact over extended behaviour, measuring both immediate and cumulative outcomes. Such experiments could clarify whether awareness has a relatively thin instantaneous functional footprint but substantial temporally extended consequences.

2. *Expand and strengthen the functional profile.* Complete systematic literature sweeps (a starting point can be found at: https://github.com/Lossfunk/consciousness-sweep) and seek converging experimental evidence across independent paradigms, including results demonstrating capacities preserved without awareness. New experiments should constrain information-strength explanations through matched first-order performance or diagnostic designs, carefully assess awareness, and distinguish genuine functional equivalence from statistically inconclusive results (Meyen et al., 2021; Michel, 2026).

3. *Refine the explanatory profile.* Investigate perceptual and affective structure, apparent presentation, and subject–world organisation through quality-space measurements, phenomenological interviews, and behavioural experiments. In particular, compare relatively unmonitored everyday experience with explicitly monitored experience, testing whether introspective elicitation, report demands, and altered states change the experience itself, access to it, or its subsequent description.

4. *Construct and intervene on candidate architectures.* Build computational agents that benefit from modelling internal processes such as attention and decision reliability. Manipulate the observability of these processes and examine whether shared or interacting internal models emerge through learning. Use ablations and other interventions to identify their functional roles, while developing explicit hypotheses about how their informational structure could account for awareness judgments, experiential organisation, and reports.

5. *Test architectures against independent constraints.* Evaluate candidates on functional and explanatory phenomena not used during their construction, alongside neural and lesion evidence. Derive prospective predictions that discriminate between alternative architectures, and test them in new tasks, altered states, clinical or developmental populations, and potentially non-human or artificial systems where appropriate. Use failures to revise both the architectures and the constraint profiles.

# Acknowledgements

ChatGPT 5.6 Sol, 6.1 Sol, and 6 Astra ("Extra High" setting) with the Codex harness were used to run literature sweeps and edit and structure parts of this paper. The authors take full responsibility for the outputs presented here.

We thank Matthias Michel for substantial discussion on a previous draft of this paper.